\documentclass[osajnl,twocolumn,showpacs,showkeys,10pt]{revtex4}  

\usepackage{graphicx}
\usepackage{dcolumn}
\usepackage{bm}

\newcommand{\ud}{\mathrm{d}}
\newcommand{\eqn}[1]{(\ref{#1})}
\newcommand{\fig}[1]{Fig.~\ref{#1}}
\newcommand{\Fig}[1]{Figure~\ref{#1}}

\begin{document}
%
\title{On the effect of dispersion on nonlinear phase noise}
\author{Keang-Po Ho}
\affiliation{Institute of Communications Engineering and Department of Electrical Engineering, National Taiwan University, Taipei 106, Taiwan.}
\email{kpho@ieee.org}

\author{Hsi-Cheng Wang}
\affiliation{Institute of Communications Engineering, National Taiwan University, Taipei 106, Taiwan.}

\date{\today}%
%
\begin{abstract}
The variance of nonlinear phase noise is analyzed by including the effect of intrachannel cross-phase modulation (IXPM)-induced nonlinear phase noise.
Consistent with Ho and Wang \cite{ho0507} but in contrary to the conclusion of both Kumar \cite{kumar05} and Green {\em et al.} \cite{green03},  the variance of nonlinear phase noise does not decrease much with the increase of chromatic dispersion.
The results are consistent with a careful reexamination of both Kumar \cite{kumar05} and Green {\em et al.} \cite{green03}.
\end{abstract}


\maketitle

Recently, phase-modulated optical communication systems are found to have wide applications in long-haul lightwave transmission systems \cite{ho05, gnauck05}.
Added directly to the signal phase, nonlinear phase noise is the major degradation for phase-modulation signals \cite{ho0507}.
However, the recent paper by Kumar \cite{kumar05} and early paper by Green {\em et al.} \cite{green03} shown that, in contrary to Ho and Wang \cite{ho0507}, nonlinear phase noise becomes much smaller for highly dispersive transmission system than that for a  system with no dispersion.
The main purpose of this paper is to reconcile the discrepancy among those three letters \cite{green03, kumar05, ho0507}.
Although there is no numerical error in both Kumar \cite{kumar05} and Green {\em et al.} \cite{green03}, the conclusion is unfortunately generalized further than its numerical results for a single pulse \cite{kumar05} or continuous-wave signal \cite{green03}.
The methods in  both  Refs.~\onlinecite{kumar05, green03}  are correct and their results are consistent largely with the results of Ref.~\onlinecite{ho0507} after a careful reinterpretation.

Ref.~\onlinecite{kumar05} found that the variance of the peak nonlinear phase noise for a single pulse decreases rapidly with chromatic dispersion.
However, a phase-modulated lightwave transmission is typically a chain of optical pulses with different modulated phases.
Ref.~\onlinecite{kumar05} ignores the nonlinear phase noise induced from adjacent optical pulses, called intrachannel cross-phase modulation (IXPM) phase noise\cite{ho0507}.
When IXPM phase noise is included, even with the model of Ref.~\onlinecite{kumar05}, nonlinear phase noise does not decrease much with chromatic dispersion.
As shown later in this letter, the IXPM phase noise may affect another pulse that is hundreds of picoseconds away from the originated pulse.

The variance of nonlinear phase noise  in Ref.~\onlinecite{green03} also decreases significantly for a continuous-wave signal with chromatic dispersion.
When the variance of nonlinear phase noise was calculated, the numerical value of Eq.~(9) of  Ref.~\onlinecite{green03} depending on the optical filter bandwidth [$\pm \Delta$ of Eq.~(9) there].
For the case of having an optical matched filter with $\Delta \rightarrow 0$ for continuous-wave signal, the variance of nonlinear phase noise of Ref.~\onlinecite{green03} is independent of the amount of chromatic dispersion [i.e., Eq.~(9) becomes Eq.~(8) there].
Another interpretation of Ref.~\onlinecite{green03} based on optical matched filter may conclude that the variance of nonlinear phase noise is independent of chromatic dispersion.
Optical matched filter is used in this letter and nonlinear phase noise does not decrease that much with chromatic dispersion, largely consistent with the results of Ref.~\onlinecite{green03} with same kind of filter.

Unfortunately, the method of Ref.~\onlinecite{ho0507} cannot directly apply to the system of both Refs.~\onlinecite{kumar05, green03} without some modifications.
Similar to the analytical results of both Refs.~\onlinecite{kumar05, gordon90}, there is an optical matched filter before the receiver.
Optical matched filter is either implicitly or explicitly assumed \cite{gordon90, kumar05, ho0507}. 
For example, the finite variance of linear phase noise [see Eq.~(25) of Ref.~\onlinecite{kumar05}] implicitly assumed a matched filter \cite{proakis4} but the numerical simulation of Ref.~\onlinecite{kumar05} assumes an ideal band-pass filter with 70-GHz bandwidth.
When white Gaussian noise with infinite bandwidth is assumed, the noise power approaches infinity.
A finite signal-to-noise ratio requires some types of optical filter and an optical matched filter has the smallest bandwidth and does not distort the signal.
Optical matched filter is used in some experimental measurements to improve the receiver sensitivity \cite{caplan01, atia99}.

To make a direct comparison with Refs.~\onlinecite{kumar05, gordon90}, we consider a transmission system consisting of two segments of equal length within an amplified fiber span.
The dispersion of the first segment is anomalous whereas that of the second segment is equal in magnitude but opposite in sign.
Within each fiber span, the accumulated dispersion as a function of distance is given by $S(z) = \beta_2 \min(z, L-z)$ where $\beta_2$ is the group-velocity dispersion coefficient and $L$ is the length of the fiber span.
In the first order, the temporal distribution of nonlinear phase noise is independent to the number of fiber spans if all fiber spans has the same configuration.

For a Gaussian pulse launching with an $1/e$-pulse width of $T_0$, at the distance of $0 \leq z \leq L$, the pulse becomes
\begin{equation}
u(z, t) = \frac{A_0 T_0}{ [T_0^2 - j S(z)]^{1/2}}
  \exp \left\{ - \frac{t^2}{2 [T_0^2 - j S(z)]} \right\}
\end{equation}
with a pulse width of $\tau(z) = [T_0^2 + S(z)^2/T_0^2]^{1/2}$, where $A_0$ is the peak amplitude.
The fiber loss is first ignored here but includes afterward.
From Refs.~\onlinecite{ho0507, ho05} and using a model similar to Ref.~\onlinecite{mecozzi00a}, the nonlinear phase noise is mainly induced by the nonlinear force of 
\begin{equation}
\Delta u_n(t) = 2j \gamma \int_{0}^{L}
   \left[ |u(z, t)|^2 n(z, t)\right]
   \otimes h_{-z}(t) e^{-\alpha z} \ud z,
\label{deltant}
\end{equation}
where $\gamma$ is the nonlinear fiber coefficient, $n(z, t)$ is the amplified-spontaneous emission (ASE) noise, $h_{-z}(t)$ is the dispersion from $z$ to $L$ with an overall dispersion of $-S(z)$, and $\alpha$ is the fiber attenuation coefficient.
The variance of \eqn{deltant} as a function of time is calculated in Ref.~\onlinecite{ho0507}.
For system without chromatic dispersion, the variance of \eqn{deltant} is equal to infinity as $n(z, t)$ is commonly assumed as white Gaussian noise.
The temporal profile of $\zeta(t) = \Delta u_n(t) \otimes h_o(t)$ is calculated here, where $h_o(t)$ is the impulse response of the optical filter preceding the receiver.
As the received signal is $s(t) = u(L, t) \otimes h_o(t)$, the nonlinear phase noise is approximately equal to 
\begin{equation}
\phi_\mathrm{nl}(t) = \frac{\Im\left\{\Delta u_n(t) \otimes h_o(t)\right\}}{s(0)},
\end{equation} 
where $\Im\{ ~\}$ denotes the imaginary part of a complex number.
Equivalently, the variance of $\phi_\mathrm{nl}(0)$ is found as the peak variance of nonlinear phase noise in Ref.~\onlinecite{kumar05}.
The variance of $\phi_\mathrm{nl}(t)$ was obtained in Ref.~\onlinecite{ho0507} for some discrete points.
Here in this letter, the whole temporal profile of the variance of $\phi_\mathrm{nl}(t)$ is derived and calculated.
If the temporal profile of $\phi_\mathrm{nl}(t)$ is concentrated around $\pm T_0$ that is the $1/e$-pulse width of $u(0, t) = u(L, t)$, the peak nonlinear phase noise from Ref.~\onlinecite{kumar05} is more than sufficient to evaluate the system performance.
However, if the temporal profile of $\phi_\mathrm{nl}(t)$ is far wider than $\pm T_0$, conclusion derived from the peak nonlinear phase noise is not sufficient to characterize the system performance.
Although the method here is similar to Refs.~\onlinecite{ho0507, ho05}, the temporal profile for the nonlinear phase noise is never shown and the discrepancy between Refs.~\onlinecite{ho0507, kumar05, green03} is never reconciled.

From the model of both Refs.~\onlinecite{kumar05, ho0507}, with prefect span-by-span dispersion compensation, the temporal profile for nonlinear phase noise is independent of the number of fiber spans.
The temporal profile for a single-span system is derived by the assumption by first-order perturbation.
With the first-order perturbation, the temporal distribution is also independent to the launched power of the signal.
For an optical matched filter with $h_o(t) = u(0, t) = u(L, t)$, we obtain $s(0) = \sqrt{\pi} A^2_0 T_0$ as the energy of each optical pulse. 
 
Using the property that both  $\Delta u_n(t)$ and $\zeta(t)$ are circular symmetric complex Gaussian random variable, after some algebra and followed Refs.~\onlinecite{ho05, ho0507}, the variance of nonlinear phase noise as a function of time is
\begin{widetext}
\begin{equation}
\sigma^2_\mathrm{nl}(t) = E\left\{ \phi_\mathrm{nl}(t)^2 \right\}
  = \frac{4 \gamma^2 A_0^2 T_0^2 \sigma_n^2}{\pi} \!\! \int_{-\infty}^{+\infty} \left| \int_{0}^L  \frac{ \exp\left\{ -\frac{t^2 - j \tau(z)^2 \omega t + S(z)^2 \omega^2 + \frac{1}{2} T_0^2 \omega^2 \left[\tau(z)^2 + 2 j S(z) \right]  }  { \tau(z)^2 - 2 j S(z) + 2 T_0^2} -\alpha z \right\}}{\sqrt{\tau(z)^2 - 2 j S(z) + 2 T_0^2}} \ud z \right|^2 \ud \omega,
\label{sigmanl}
\end{equation}
\end{widetext}
\noindent where $2\sigma_n^2$ the power spectral density of ASE noise at the launching location of $z = 0$.

\begin{figure}
\centerline{
\includegraphics[width = 0.49 \textwidth]{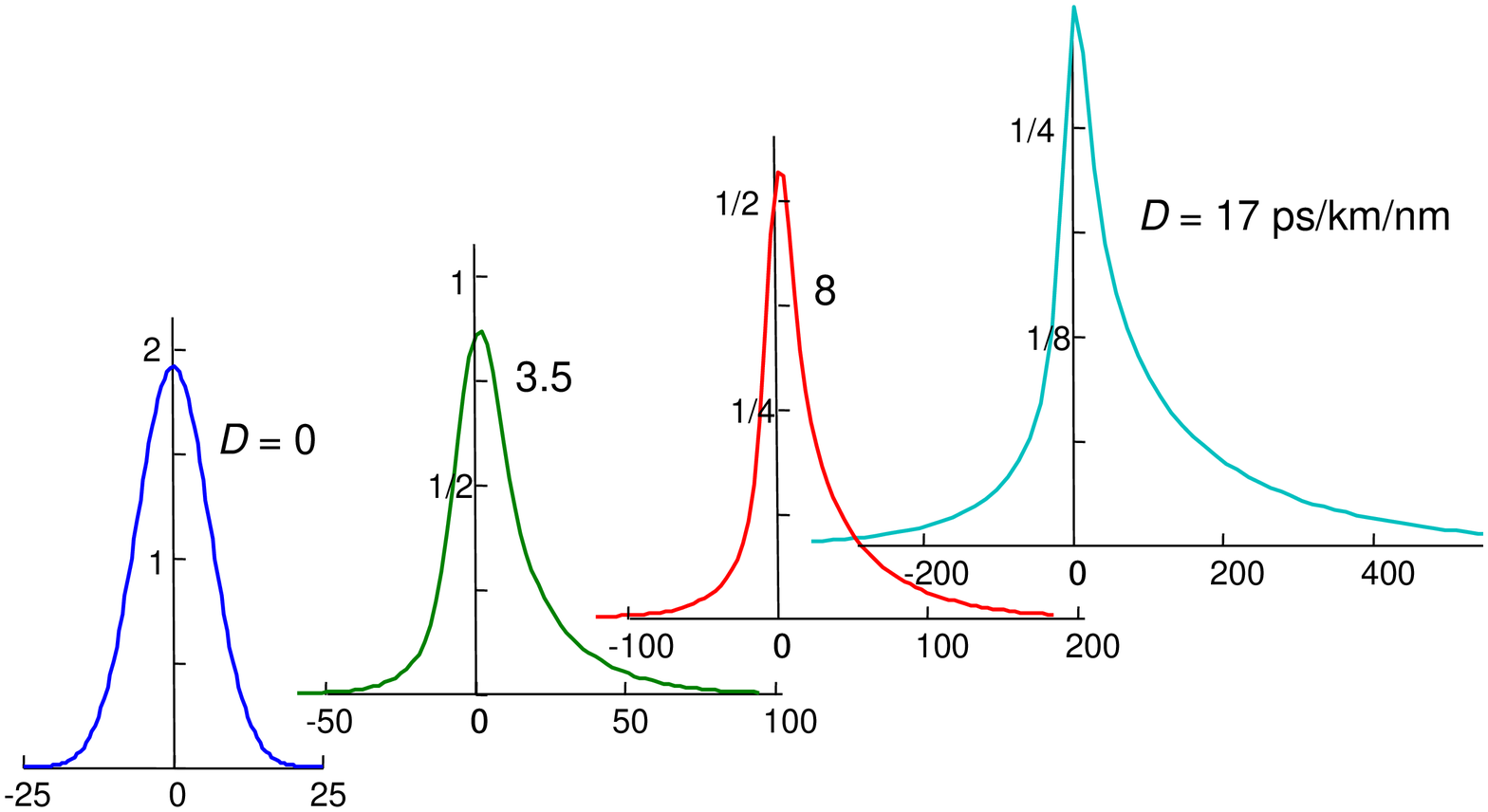}}
\caption{The temporal distribution of the STD of nonlinear phase noise $\sigma_\mathrm{nl}(t)$.
The $x$-axis is time in picosecond and the $y$-axis is $\sigma_\mathrm{nl}(t)$ in arbitrary linear unit.
Note that the $x$- and $y$-axes with difference dispersion do not have the same unit.
}
\label{figprofile}
\end{figure}

\Fig{figprofile} shows the temporal profile of the standard deviation (STD) of nonlinear phase noise of $\sigma_\mathrm{nl}(t)$ as a function of time.
The $y$-axis of \fig{figprofile} is in arbitrary linear unit.
\Fig{figprofile} is calculated for a 80-km fiber link with dispersion coefficient of $D = 0, 3.5, 8,$ and $17$ ps/km/nm and an initial launched pulse of $T_0 = 5$ ps.
As explained earlier, the temporal distribution by itself is independent of the number of fiber spans.  
The number of fiber spans, signal-to-noise ratio, and launched power scale the whole curves of \fig{figprofile} up or down.

The $x$-axis of \fig{figprofile} does not have the same scale.
The $x$-axis of the curve in \fig{figprofile} having a larger dispersion coefficient is scaled down by a factor of 2 than the one with smaller dispersion coefficient.
Similarly,  the $y$-axis of the curve in \fig{figprofile} having a larger dispersion coefficient is scaled up by the same factor of 2 than the one with smaller dispersion coefficient.
After the scaling, all curves in \fig{figprofile} have more or less the same height and width.

\Fig{figprofile} confirms the conclusion of Ref.~\onlinecite{kumar05} that the peak nonlinear phase noise decreases rapidly with chromatic dispersion.
In term of STD, the peak nonlinear phase noise with a chromatic dispersion of $D = 17$ ps/km/nm is about 7 times less than the dispersionless case of $D = 0$.
However, the temporal distribution of the nonlinear phase noise also broadens rapidly with chromatic dispersion.
With a range from $-200$ to $+400$ ps, the temporal distribution of $\sigma_\mathrm{nl}(t)$ for $D = 17$ ps/km/nm is about 20 times wider than the case without dispersion of $D = 0$ of within $\pm 15$ ps.

If the effect of chromatic dispersion to ASE noise is ignored (the model of Ref.~\onlinecite{kumar05}),  IXPM phase noise from adjacent pulses to the same pulse is 100\% correlated.
If the effect of chromatic dispersion to ASE noise is included, the correlation between IXPM phase noise from adjacent pulses decreases slightly but the tail of the temporal profile increases.
For a qualitative understanding without repeating the calculations in Refs.~\onlinecite{ho0507, ho05}, we can assume that the nonlinear phase noise induced by adjacent pulses to the same pulse is highly correlated.
For highly correlated noise, the combined noise has a STD approximately equal to the sum of the individual STD, approximately the same as the area of the curves of \fig{figprofile}.
Because main parts of the four curves in \fig{figprofile} have the same peaks and width after scaling up in height and down in time by the same factor, nonlinear phase noise does not decrease that much with the increase of chromatic dispersion \cite{ho0507, ho05}.
 
Other than the dispersionless case with $D = 0$, the temporal profile of nonlinear phase noise is asymmetric with respect to the original center of the pulse of $t = 0$.
In Ref. \onlinecite{green03}, a shift was observed in frequency domain but the shift in time domain of \fig{figprofile} is first observed in this letter.
The peak of nonlinear phase noise is located approximately at the center of the pulse but shifted slightly to positive time.
The asymmetric temporal profile is due to the inclusion of the dispersive effects to the ASE noise.
Without the inclusion of the dispersive ASE, the temporal profile of $\sigma_\mathrm{nl}(t)$ is symmetrical with respect to the pulse center of $t = 0$.

The model here is very similar to the model of Ref.~\onlinecite{kumar05}.
Only the first-order term is used here but Ref.~\onlinecite{kumar05} also included a minor second-order term [see Eq.~(16) there].
The model here includes the correlation of $E \left\{ n(z_1, t + \tau) n(z_2, t) \right\}$ with a power spectral density of $2\sigma_n^2 \exp\left\{ j [S(z_1) - S(z_2)] \omega^2/2 \right\}$.
The correlation of ASE noise is ignored in Ref.~\onlinecite{kumar05}.
The temporal profile here is asymmetric but the temporal profile of Ref.~\onlinecite{kumar05} [given by $h_r(t)$ in Eq.~(16) there] is symmetrical.
The temporal profile $h(t)$ in Ref.~\onlinecite{kumar05} [Eq.~(16) there] is very similar to \eqn{sigmanl} here.
If IXPM phase noise is included to the numerical method of Ref.~\onlinecite{kumar05}, different conclusion should be arrived.

The optical matched filter for continuous-wave optical signal is a very narrow-band optical filter.
Using a very narrow-band filter in the model of Ref.~\onlinecite{green03}, the nonlinear phase noise is independent of chromatic dispersion there.
If the optimal optical filter is used to detect a signal, both Refs.~\onlinecite{green03, ho0507} should arrive with similar results.

This letter finds that all three letters of Refs.~\onlinecite{green03, kumar05, ho0507} should provide consistent results if IXPM phase noise is included in Ref.~\onlinecite{kumar05} and optical matched filter is used in Ref.~\onlinecite{green03}.
Nonlinear phase noise does not decrease much with the chromatic dispersion in a practical lightwave transmission system.
With optical matched filter and according to both Refs.~\onlinecite{green03} and \onlinecite{ho0507}, the nonlinear phase noise for system with large dispersion is approximately equal to an equivalent dispersionless continuous-wave system having the same power.

If the correlation of ASE noise due to chromatic dispersion is included to the model, the temporal profile of the STD of nonlinear phase noise is asymmetrical with respect to its peak.
The time-domain asymmetric profile is first observed for nonlinear phase noise here.


\end{document}